\RequirePackage{fix-cm}
\RequirePackage{multirow}
\documentclass[smallextended]{svjour3}    
\PassOptionsToPackage{numbers}{natbib}
\usepackage{natbib}
\usepackage{wrapfig}
\usepackage{lineno,hyperref}
\usepackage[english]{babel}
\usepackage{footnote}
\makesavenoteenv{tabular}
\makesavenoteenv{table}
\usepackage{booktabs}
\usepackage{csvsimple}
\usepackage{pdflscape}
\usepackage[autolanguage]{numprint}
\usepackage{pgfplots,pgfplotstable}
\usepgfplotslibrary{statistics}
\usepackage[textsize=footnotesize,backgroundcolor=orange!20,linecolor=orange,bordercolor=orange,textwidth=12em]{todonotes}
\usepackage{caption}
\usepackage{subcaption}
\usepackage{placeins}
\usepackage{framed}
\usepackage{rotating}
\usepackage{multirow}
\usepackage{tabularx}
\usepackage{tablefootnote}
\usepackage{graphicx}
\usepackage{tikz}
\usepackage{collcell}
 
\begin{document}

\title{Architecture Smells vs. Concurrency Bugs: an Exploratory Study and Negative Results}


\author{Damian Andrew Tamburri \and Francesca Arcelli Fontana \and Riccardo Roveda \and Valentina Lenarduzzi}

\institute{Damian A. Tamburri \at TUe - JADS, 's-Hertogenbosch, Netherlands \\
\email{d.a.tamburri@tue.nl}
\and
Francesca Arcelli Fontana \at University of Milano - Bicocca, Milano, Italy \\
\email{arcelli@disco.unimib.it} 
\and
Riccardo Roveda \at Alten Italia, Milano, Italy \\
\email{riccardo.roveda@alten.it}
\and 
Valentina Lenarduzzi (corresponding author) \at University of Oulu, Finland \\
\email{valentina.lenarduzzi@lut.fi}   
}

\maketitle

\begin{abstract}
Technical debt occurs in many different forms across software artifacts. One such form is connected to software architectures where debt emerges in the form of structural anti-patterns across architecture elements, namely, \emph{architecture smells}. 
As defined in the literature, ``Architecture smells are recurrent architectural decisions that negatively impact internal system quality", thus increasing technical debt.
In this paper, we aim at exploring whether there exist manifestations of architectural technical debt beyond decreased code or architectural quality, namely, whether there is a relation between architecture smells (which primarily reflect structural characteristics) and the occurrence of concurrency bugs (which primarily manifest at runtime).
We study 125 releases of 5 large data-intensive software systems to reveal that (1) several architecture smells may in fact indicate the presence of concurrency problems likely to manifest at runtime but (2) smells are not correlated with concurrency in general --- rather, for specific concurrency bugs they must be combined with an accompanying articulation of specific project characteristics such as project distribution. 
{\color{black}As an example, a cyclic dependency could be present in the code, but the specific execution-flow could be never executed at runtime.}

\end{abstract}

\section{Introduction}\label{sec:introduction}

%
%
%
%
Software architecture is a fundamental abstraction of software systems, their visible properties, and operating parts \cite{BCK98}. In this paper we investigate the existence of a relation between issues in the architecture and \emph{concurrency bugs}, i.e., issues that are intimately tied to architectural behavior as opposed to code flaws (e.g., code smells \cite{PalombaZFLO16}). This investigation is impactful for several reasons. First technical debt (TD) may be used to study concurrency bugs from a different perspective; the architectural technical debt metaphor offers such a perspective by summarizing the technical compromises made on software artefacts (code, models, architecture, and more) that, on the one hand, may yield short-term benefits but, on the other hand, reveal themselves as long-term technical investments accumulating additional project costs for their resolution~\cite{AvgeriouKOS16} and which may lead to nasty and unforeseen runtime manifestations whose costs of occurrence are not clear nor managed in any way. 

Similarly, of the many afflictions of TD, software architecture is considerably important, since it summarizes the entire system and is often the first artifact to be used for TD resolution~\cite{koziolek2013}, but how that impacts concurrency dimensions of the system remains unknown. What is more, compromising architecture elements and their structural dept characteristics may have unforeseen negative impacts many non-functional characteristics beyond concurrency levels, e.g., performance, reliability, and maintainability of the entire application; a dept-resolved lens over these may help further maintenance and evolution practices and their prioritization. 


To address the aforementioned gap, in this study we focused our attention on understanding the degree to which known sub-optimal architecture circumstances---architecture \emph{smells} \cite{Abebe2009,BCK98}---impact the concurrency characteristics of data-intensive software systems. More in particular, we aim at establishing whether there exist architectural violations that reflect a higher chance of generating concurrency bugs. 

To conduct our empirical inquiry, we use Arcan~\cite{ArcelliFontana2016c}, a tool developed to detect several major architectural smells (AS), structural indicators that architectural TD may be present. 
We exercised Arcan over 125 releases of 5 large, open-source software packages. We overlapped the concurrency bugs coded independently~\cite{AsadollahSEH17} in our dataset with Arcan smells detections.

{\color{black}If there exists relation between architectural smells and concurrency bugs then, it will be possible to avoid concurrency bugs by not introducing or removing the architectural smells.
We expect to find correlations between different architectural smells and concurrency bugs. As example, we expect to find a good correlation between cyclic dependency and deadlock.}

Our results indicate that architecture smells reflect concurrency bugs only if the projects under analysis exhibit two characteristics at the same time, that is: (1) they are intrinsically distributed, meaning that they are designed to support distributed processing and are themselves parallel systems ---this is likely because in distributed systems the negative influence of architecture smells becomes more grievous due to parallelization, which could be established through further experimentation; (2) they are updated and released very often---this is likely because of the frequency of code changes required by architecture smells which necessarily leads to making more concurrency-related issues while coding a distributed system.

This manuscript presents 2 major contributions: 


\begin{enumerate}
\item The exploration of empirical relations between architecture smells and concurrency bugs. This contribution impacts research and practice in several ways, e.g., by offering an architecture-specific lens to concurrency bug prediction.
\item The subsequent evaluation of Arcan~\cite{ArcelliFontana2016,ArcelliFontana2016c} in action; Arcan is a tool for automated architecture smells detection. 
\end{enumerate}

\textbf{Structure of the Paper}. The rest of this paper is structured as follows. First, Sec. \ref{sec:related}
outlines related work.  Sec.~\ref{sec:arcan} outlines
our empirical study design. Finally, Sec.~\ref{sec:results} and Sec.~\ref{sec:conc} outline evaluation results and conclude the paper, respectively.

\section{Related Work}\label{sec:related}
Several works offer a motivational basis or background for our work.

\subsection{Architectural Smells Detection}\label{sec:detect}

Many tools have been developed for code smells detection but only a few tools are currently available for architectural smells detection \cite{Azadi19}. The following briefly reports on some of them. 
AiReviewer\footnote{\url{http://www.aireviewer.com}}
is a commercial tool that supports the detection of both code smells and some
design or architectural smells. 
 Designite\footnote{\url{http://www.designite-tools.com}} is another commercial tool able to detect several design smells in C\# and Java projects. 
 Arcade \cite{Arcade} is a tool  developed for architectural smell detection in Java projects, the authors propose also a classification of the smells.
 DV8~\cite{cai2019dv8} is a commercial tool able to detect six architectural smells, called anti-patterns~\cite{MoWicsa2015}, five patterns defined at the file level and one at the package level.
Other commercial tools are Sotograph, Sonargraph, Structure 101 and Cast, which are not specialised in architectural smell detection, but are able to detect different kinds of architectural violations, such as dependency cycles. 

Other tool prototypes have been proposed, e.g., SCOOP~\cite{MaciaICSM2012}, and one from Garcia et al.~\cite{Garcia-ASE-2011}. 
Moreover, another tool to measure software modularity, and to detect architecture anti-patterns were defined by Cai and Kazman~\cite{Cai2019}.
Finally,  Arcan is the tool we used in this study able to detect  different  architectural smells through the analysis of compiled Java files. Through Arcan it is possible to inspect the results through graphs, more useful
respect to other views, allowing  to better identify refactoring opportunities for the architectural smells.
 The tool is freely available on request at \url{http://essere.disco.unimib.it/wiki/arcan}.
 

\subsection{Empirical studies on Architectural Issues Evaluation}\label{sec:smells} 
In this section we report key empirical studies done or related to architectural issues and smells. 

Le et al. \cite{Le2016} propose an approach to build bug prediction models based on both implicit and explicit features of a software system.  The authors especially  consider architectural-based features, with a focus on architectural smells and architectural decay metrics.

According to architectural changes, Behnamghader et al.~\cite{Behnamghader2017} conduct a large empirical study of changes found in software architectures spanning many versions of different open-source systems. Their study reveals several new findings regarding the frequency of architectural changes in software systems, and the common points of departure in a system architecture during  system's maintenance and evolution.

Moreover, some works explore the correlations between code smells or code anomalies  and architectural smells. 

Macia et al.~\cite{Macia2012b} studied the relationships between code anomalies and architectural smells in 6 software systems (40 versions).
They considered 5 architectural smells  and 9 code smells.
They empirically found that each architectural problem represented by each AS often reflects multiple code anomalies. More than 80\% of architectural problems were  related to code anomalies.
Oizumi et al.~\cite{Oizumi2014} analyzed 7 systems and suggest that  certain topologies of code smells agglomerations are better indicators, than others, of
architectural problems. They have considered six code smells detected through the rule of Lanza Marinescu~\cite{Lanza2006} and 7 architectural smells detected through rules defined by Macia.
Arcelli Fontana et al~\cite{ArcelliJSS2019} investigated the correlation between architectural and code smells founding no statistical significant results.

We have not found in the literature any works on the analysis of the correlation between architectural smells and concurrency bugs --- the key goal behind the work in this manuscript.

\subsection{Concurrency Bugs in Data-Intensive Systems Evaluation}

According to the taxonomy in Tchamgoue et al. \cite{978-3-642-27207-3_48} there exist four classes of concurrency bugs, namely: (1) data races ---that is, a shared memory is uncoordinately accessed with at least one write by both an event handler and a non-event code or by multiple event handlers; (2) atomicity and order violations ---that is, the conditions when a program fails to enforce the programmers' execution order intention; (3) deadlocks - that is, when an instruction or interleaving contains a blocking operation that blocks indefinitely the execution of a program; (4) stack overflow when a program attempts to use more memory space than is available on the stack. 
Several works have touched upon all the afore-mentioned classes of concurrency bugs, their nature, and nurture as well as their nasty effects on parallel and concurrent systems' operation. But little is known still on the architectural choices or code smells which are more prone to concurrency violations. 

In the concurrency bugs detection side, Zhang et al. \cite{Zhang} and later Kidd et al. \cite{KiddRDV11} propose tools specifically designed to address specific classes of concurrency bugs, but their and similar approaches are limited to detection and avoidance in the context of operative systems design and operation rather than their maintenance and evolution. Also, similar works tend to offer domain-specific lens of analysis with limited generalisability, e.g., the works by Park \cite{Park12,Park13}. 

More in line with maintenance, the work by Kelly \cite{KellyWLM09} predicates the successful use of control engineering and control theory with the goal of limiting the emergence and ultimately eradicating concurrency bugs, at the same time, however, their work is limited in that it looks at very low-level abstractions (i.e., programming constructs, bytecode, etc.) rather than providing an investigation that serves higher levels of abstraction such as our own (architecture level). Furthermore, along the lines of maintenance by testing and refactoring, Smith \cite{Smith14} and later Fonseca \cite{fonseca2015effective} cover an extensive study of concurrency bugs diagnosis and removal, theoretically assessing their link with architecture-level constructs and patterns as well as proposing early support for automated refactoring of concurrency bugs.  Similarly to these works, Kelk et al. \cite{KelkJB13} offer a first, very specific attempt at automated program repair designed to address specific classes of concurrency bugs. From a more generalist perspective, Thompson explores the notion of \emph{mechanical sympathy}, a programming model wherefore the increasingly required synergy between hardware and software makes the investigation of the issues which reflect on both of paramount importance.

In line with these works, we choose to further the understanding and state of the art in software architecture research as well as architecture-driven evolution and modernization of concurrent systems by looking into the relation between concurrency bugs and architecture smells in the context of distributed systems such as Big Data middleware. 

Finally, from a concurrency-specific perspective, the state of the art has concentrated on assessing the recurrence of new types of concurrency bugs (e.g., Accuracy Bugs \cite{AkturkAIMK16})  and their impact on distributed systems' quality properties. Such works reinforce the need of further understanding into maintaining large-scale distributed system architectures with a bug- and concurrency-centric lens, e.g., with increasing granularity level. In the scope of this study we intended to pursue this research direction starting from the first possible level of abstraction, i.e., considering concurrency bugs regardless of their intimate nature (e.g., \emph{intra-} vs. \emph{inter-}process concurrency bugs or similar distinctions).

\subsection{Summary and Study Context}

From the overview above it is clear that there is an increasing need to study software architectures in the context of distributed systems to understand how concurrency bugs are produced and addressed. On one hand, the techniques currently available mostly take an event-driven approach or focus on specific classes of bugs, avoiding to touch higher-order artefacts such as software architectures, which, de-facto, drive the actual development of the system. On the other hand, software architecture and design constructs (e.g., software architecture and design metrics, design patterns, code smells, architecture smells, etc.) can aid in the detection and avoidance of classic concurrency bugs, especially within the context of emerging concurrent and distributed systems where reactive-programming \cite{RamsonH17} and reactive architecture styles \cite{DebskiSMSM18} are increasingly being considered in both research and practice \cite{voellmy2010}.

In line with the above limitations, we focus on representatives of both \emph{intra-} (i.e., within the scope of a process) and \emph{inter-}process (i.e., between multiple processes) concurrency bugs, to assess if their occurrence is in any way related to the emergence of software architecture smells.


\section{Study Design}\label{sec:arcan}

\subsection{Research Problem and Question}

In the scope of this study, we focus on identifying and \emph{measuring} the overlaps and correspondences between  architecture smells and concurrency bugs  (defined in section~\ref{sec:arcan}) whose diagnosis and repair has not yet found an exhaustive solution~\cite{KhoshnoodKW15}. The \emph{focus} and \emph{context} of concurrency is well justified by the recent emergence of highly concurrent and decentralized systems, e.g., based often on microservice architectures. In the scope of such advanced architecture styles, it becomes critical to provide to the \emph{perspective} of software architects, ways to understand and quantify the impact (if any) of known software architecture issues over the emergence and risk of concurrency bugs. Finding any relation thereof, would essentially provide more advanced design principles to structure highly concurrent and distributed systems.

Hence, we aim to answer the following research questions:

\vspace{2mm}
\noindent\textbf{RQ$_1$.} To what extent do known architectural smells impact on the emergence of major concurrency bugs?

\vspace{2mm}
\noindent\textbf{RQ$_2$.} Is the presence of architectural smells independent from the presence of concurrency bugs?

\subsection{Architectural Smells Analysis}\label{asmells}

We describe below the architectural smells (AS) we have considered in this study:
 
 \begin{enumerate}

  \item  \emph{Unstable Dependency (UD)}: describes a subsystem (component)
  that depends on other subsystems that are less stable than itself. 
  This may cause a ripple effect of changes in the system~\cite{Marinescu2012}. Detected on packages.
   \item\emph{Hub-Like Dependency (HL)}: arises when an  abstraction has (outgoing and ingoing) dependencies with a large number of other abstractions~\cite{Suryanarayana2015}. Detected on  classes and packages.
  \item  \emph{Cyclic Dependency (CD)}: refers to a subsystem (component) that is involved in a chain of relations that break the desirable acyclic nature of a subsystem's dependency structure. The subsystems involved in a dependency cycle can be hardly released, maintained or reused in isolation. Detected  on classes and packages.
  Arcan is able  to detect the cycles according to  different shapes (see Figure 1) as described in ~\cite{DBLP:conf/aswec/Al-MutawaDMM14}.
  \item  \emph{Multiple Architectural Smell (MAS)}: this is not actually a smell, but we list it here for simplicity. It is an index useful to identifies a subsystem 
  (component) that is affected by more than one architectural smell and
  provides the number of the architectural smells involved. 

\end{enumerate}

 
 We decided to consider these AS in the study since they represent relevant  problems related to dependency issues: components highly coupled and with a high number of dependencies cost more to maintain.
Hence, they can be considered more critical and can lead to a progressive architectural debt \cite{Ernst2015}. In particular, Cyclic Dependency is one of the most common architectural smell, more dangerous and difficult to remove \cite{Martini2018}, \cite{Rizzi2018}.

AS have been detected through the Arcan tool~\cite{ArcelliFontana2016c},~\cite{Arcelli2017-ICSA},
The tool relies on graph database technology {\color{black} to perform graph queries, which let on higher scalability during the detection process and management of a huge number of different kinds of dependencies}.
Once a Java project has been analyzed by Arcan, a new graph-database is created containing the structural dependencies of the projects. It is then possible to run detection algorithms on this graph to extract information about the analyzed project: package/class metrics and architectural issues.

{\color{black} This tool detects architectural smells based on dependency issues, by reckoning different metrics proposed by Martin~\cite{Martin1995} such as those related to instability issues}. 
The evaluation of Arcan detection performances, through standard Information Retrieval performance metrics,(i.e., confusion matrix elements and derivatives, like precision and recall) is reported in \cite{Arcelli2017-ICSA} and an evaluation on 10 open source projects in \cite{ArcelliFontana2016c}.
The results of Arcan were validated using the feedbacks of the practitioners on four industrial projects \cite{Martini2018}.
{\color{black} Moreover, another recent study~\cite{ArcelliJSS2019} applied Arcan to detect architectural smells in 102 projects from Qualitas Corpus~\cite{Terra:2012}.}
The tool is freely available and easy to install and to use.
~\footnote{http://essere.disco.unimib.it/wiki/arcan}.


\begin{figure}[tb]
\centering
\includegraphics[width=0.8\linewidth,clip=true,trim=300 220 320 220]{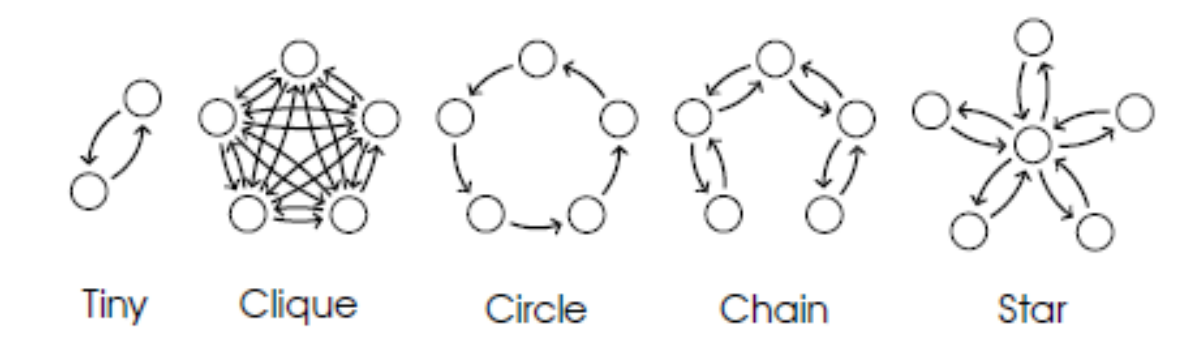}
\caption{Cycles shapes~\cite{DBLP:conf/aswec/Al-MutawaDMM14} in the scope of architecture technical debt detection.}
\label{cycleShapes}
\end{figure}

\subsection{Concurrency Bugs }\label{sec:emp}


We describe below a number of typically-occurring \cite{ZhouNG15,Leesatapornwongsa16} concurrency bugs that we have considered in our study.

\begin{enumerate}
\item \emph{\textbf{Data-Race (DR):}} DR occurs when at least two threads access the same data and at least one of them write the data~\cite{yoshiura2014static}. The nasty effects of data-racing occur when concurrent threads perform conflicting accesses by trying to update the same memory location or shared variable~\cite{henningsson2004assuring,akhter2006multi}.
\item \emph{\textbf{Deadlock (DL):}} DL is a condition in a system where a process cannot proceed because it needs to obtain a resource held by another process but it itself is holding a resource that the other process needs~\cite{bhatia2014deadlocks}. More generally, it occurs when two or more threads attempts to access shared resources held by other threads, and none of the threads can give them up ~\cite{henningsson2004assuring}. During deadlock, all involved threads are in a waiting state. 
\item \emph{\textbf{Livelock (LL):}} LL happens when a thread is waiting for a resource that will never become available while the CPU is busy releasing and acquiring the shared resource. It is similar to Deadlock except that the state of the process involved in the livelock constantly changes and is frequently executing without making progress~\cite{chapman2008using}.
\item \emph{\textbf{Starvation (SS):}} SS is a condition in which a process indefinitely delayed because other processes are always given preferences~\cite{stallings2012operating}. Starvation typically occurs when high priority threads are monopolizing the CPU resources. During starvation, at least one of the involved threads remains in the ready queue.
\item \emph{\textbf{Suspension-based locking (SBL):}} SBL occurs when a calling thread waits for an unacceptably long time in a queue to acquire a lock for accessing a shared resource~\cite{lin2013supporting}.
\item \emph{\textbf{Order violation (OV):}} OV is defined as the violation of the desired order between at least two memory accesses~\cite{jayasinghe2010core}. It occurs when the expected order of interleaving does not appear~\cite{park2010falcon}. If a program fails to enforce the programmer intended order of execution then an Order violation bug could happen~\cite{lu2008learning}.
\item \emph{\textbf{Atomicity violation (AV):}} AV refers to the situation when the execution of two code blocks (sequences of statements protected by lock, transactions, etc.) in one thread is concurrently overlapping with the execution of one or more code blocks of other threads such a way that the result is inconsistent with any execution where the blocks of the first thread are executed without being overlapping with any other code block.
\end{enumerate}

\subsection{Architecture Smells as mediators for Concurrency Bugs}

In the following we discuss any connections and overlaps we report from the state of the art concerning the concurrency bugs under investigation and formulate a series of theoretical conjectures to be tested out in the scope of our empirical study. 

\paragraph{Data-Race}
Architect Michael Barr explains\footnote{\scriptsize\url{https://barrgroup.com/Embedded-Systems/How-To/Top-Ten-Nasty-Firmware-Bugs}} that Starvation is connected to any condition wherefore the architectural structure does not warrant for proper Deadlock and racing freedom (i.e., riddance of data race conditions). 

In the scope of our work, we aim to show whether reported race-conditions for a complex software architecture may also be connected to its dependency structure, and any smells thereof. For example, Demsky et Al. \cite{Demsky:2010:VOC:1806799.1806858} outline an architectural refactoring tactic where software architecture views are used to instrument fine-grained locking mechanisms that visibly reduce or altogether avoid data-race conditions. The tactic in question would work by reducing the number of dependencies across software architecture modules and refactoring ``gateway'' modules that act as viewers to supervise architecture operations.


\paragraph{Deadlock}

The Microsoft Development Handbook identifies\footnote{\scriptsize\url{https://technet.microsoft.com/en-us/library/ms177433(v=sql.105).aspx}} recurrent deadlock scenarios as being connected to unstable dependency scenarios wherefore packages and modules depend upon packages and modules that are less stable than them. Literature generally agrees with the previous statement, remarking how circularity and hierarchical structures offer fertile grounds for deadlocks to thrive \cite{Muhanna91}. For example, in distributed deadlock conditions, the modularity structure and the degree of distribution of software systems encourage deadlocks to manifest \cite{KshemkalyaniS94}. 


\paragraph{Livelock}
Livelock happens when a thread is waiting for a resource that will never become available while the CPU is busy releasing and acquiring a shared resource currently busy elsewhere in the software architecture. The Livelock condition is similar to deadlock except that the state of the process involved constantly changes and is frequently executing without making progress~\cite{chapman2008using}. Similarly to deadlock, we expect livelocks to be connected to an increasing number of mutual dependencies present across the architecture --- the more complex the dependency structure, the more likely is that deadlocks would occur in their active state, i.e., livelocks~\cite{DobricaN02}. However, the existence of livelocks requires an additional element of active mutual execution between the busy, resource-occupying process and the active, resource-requiring process. this additional element is likely to make the occurrence of Livelock concurrency bugs lesser than the generic deadlock counterpart. 

\paragraph{Starvation}

Starvation is ``a condition in which a process indefinitely delayed because other processes are always given preferences?"~\cite{stallings2012operating}. Starvation typically occurs when high-priority threads are monopolizing the CPU resources. During starvation, at least one of the involved threads remains in the ready-queue. On one hand, Architect Michael Barr et Al. explain that Starvation is connected to any condition wherefore the architectural structure does not warrant for proper racing freedom (i.e., avoidance of data races), a condition which is well-outlined by en-field investigations in Software Performance Engineering (SPE) and its study of software architectures. On the other hand, SPE researchers have long been studying the connection between high-density component-based designs and their proneness to race conditions, e.g., in the context of real-time~\cite{Zalewski05} or mobile systems \cite{GrassiC00}. 

\paragraph{Suspension-based locking}

Often referred to as ``Spinlock", SBL is a lock which causes a thread trying to acquire it to simply wait in a suspend loop (``spin") while repeatedly checking if the lock is available. By its own definition, SBL is a condition where a highly decoupled architecture likely features interdependent modules which require more appropriate or specific synchronization mechanisms \cite{Schutz01,KostrzewaTSE16}. 


\paragraph{Order and Atomicity violation}

Both order and atomicity violation reflect internal behavioral characteristics of modular structures and architecture elements. We were not able to formulate any conjectures reflecting these two concurrency bugs ourselves, however, literature on the matter seems to concur that several structural peculiarities and patterns across complex software architectures lead to an increase in buggy conditions~\cite{LuPZ12,LetkoVK08,WuYC15}. For example, Shangru et Al.~\cite{WuYC15} adopt an abstraction sub-space reduction approach to identify sub-spaces in the variable abstraction space across systems execution traces which may reflect buggy conditions thus aiding in the removal of reported violations. 
In much the same vein, we argue that Arcan helps uncovering architectural implications (if any) around violation bugs in complex software architectures - these implications likely exhibit a recurrent, time-resolved presence across a number of projects and across a number of their releases.

Time-resolution (i.e., recurrence of correlation over several subsequent releases~\cite{Shull2008}) is required for an architecture smell to be effectively considered a proxy of violation bugs. Also, for the sake of generalization, such time-resolution needs to manifest across a statistically relevant sub-set of projects across our sample~\cite{Wohlin06}.

\subsection{Data Sampling and Control Factors}\label{sec:variables}
The principal research question in our empirical study assumes the availability of an independently-coded evaluation
dataset can be used, featuring projects with architecture issues made
explicit by an independent party. As a key requirement for our dataset to fit with our purpose, the projects involved need to be intrinsically distributed and show evidence of a parallel computing paradigm in their programming model. As such, our dataset was constructed focusing solely on large-scale Big Data middleware which are distributed \emph{by-design}. With this latter explicit choice we aimed at control the \emph{distributiveness} factor by construction with our sampling strategy --- more specifically, all the projects we selected in our sample are distributed both in terms of applications' operations (i.e., all projects are data-intensive middleware supporting the operations of distributed big-data processing applications) and in terms of platform behavior (i.e., the platforms themselves are intrinsically distributed systems).

However, for the sake of generalization, the dataset in question, needs to reflect a series of controllable variables, in order to offer a reliable
evaluation. In our case, we chose to focus on controlling the following variables to obtain a viable dataset:

\begin{itemize}
\item \textbf{\emph{Project Size:}} The presence of certain architecture smells could be higher in projects with certain size; for this reason we decided to consider equally both small/medium (200 KSLOC - 500 KSLOC) and large ($>$ 501 KSLOC) projects.

\item \textbf{\emph{Team Size:}} The presence of certain organizational dynamics could lead to the emergence of specific architecture smells; we aimed at assessing the validity of Arcan in several team circumstances wherefore technical debt detection makes sense. In this respect, we ensured team sizes across our dataset, to warrant a sufficient coverage of small ($<$25 participants), medium (20-30 participants) and large ($>$35 participants) communities.

\item \textbf{\emph{Project Popularity:}} Popular projects tend to be subject of continuous releases and continuous refactoring to code; these circumstances may therefore increase the number of architectural smells across the sample such that Arcan correctness in detection may be compromised. In this respect, we controlled for project popularity making sure that a sufficiently different number of stargazers (i.e., the number of stars the project has received from how many distinct users) are present for the projects we consider in our dataset.

\item \textbf{\emph{Project Type:}} Very complex products and projects may tend to be more error-prone at the architectural level than simpler projects; these circumstances may compromise the ability of Arcan to detect architecture smells where code or architectural complexity rather than code size may be compromising project features. In this respect we controlled that the projects in our sample are equally distributed across 5 categories: (1) Middleware, (2) Software Application, (3) Development Library, (4) Application Framework and (5) Scheduling System. All 5 characteristics are non-overlapping, meaning that they reflect either a different level of architectural abstraction, a different application lifecycle phase, or a different target component type (i.e., the type of architectural component that the project is aimed at supporting).

\item \textbf{\emph{Bug Quality/Quantity:}} Projects whose releases contain more bugs of a certain type may invalidate our study because of an uneven probability distribution over the entire dataset of the individual bug types. We chose to provide an equilibrium of bug types, sampling projects and releases such that an equal proportion of each individual concurrency bug type with respect to the sizes of involved projects.
\end{itemize}

To address the above data requirements, we investigated the literature and open datasets available from several communities (e.g., Mining Software Repositories, International Conference on Software Maintainance and Evolution to name a few). Our choice fell on the dataset made available by Asadollah et al. in~\cite{AsadollahSEH17}. The authors analyze over 125 releases evenly arranged across 5 open-source projects, namely: Apache Hadoop, Apache ZooKeeper, Apache Oozie, Apache Accumulo, and Apache Spark. The projects are recapped on Table~\ref{tab:projects} - the table shows the projects' demographics gathered from \url{https://www.openhub.net}, in particular, rows address the controllable variables of the study described previously, while the columns name the specific project.

\begin{table}[htbp]
\centering
\caption{Projects demographic data of the dataset}
\resizebox{\linewidth}{!}{
    \begin{tabular}{r|p{1.4cm}p{1.2cm}p{1.1cm}p{1.2cm}p{0.5cm}} \hline 
    & \multicolumn{5}{c}{\textbf{Apache Projects}}\\ \hline 
    & \textbf{Accumulo} & \textbf{Hadoop} & \textbf{Oozie}  & \textbf{Spark} & \textbf{ZooKeeper} \\
    \textbf{Releases} & 11    & 76    & 6     & 12     & 20 \\
    \textbf{Size}  & 367.854 & 2.434.336 & 194.599 & 1.207.823 & 144.322 \\
    \textbf{Popularity} & 24    & 117   & 1   & 50    & 22 \\
    \textbf{Team Size} & 94     & 210   & 29  & 1552  & 18 \\
    \textbf{Type} & (3)  & (1) & (5) & (2) & (4) \\
    \textbf{Activity} & H
    & VH 
    & M
    & VH 
    & VL \\ \hline 
    \multicolumn{6}{l}{\textbf{Legenda -}  \textit{Project Type:} (1) Middleware; (2) Software Application; }\\
    \multicolumn{6}{l}{(3) Development Library; (4) Application Framework; (5) Scheduling System.}\\
    \multicolumn{6}{l}{VH means Very High, H means High, M means Moderate, and VL means Very Low}
    \end{tabular}
    }
  \label{tab:projects}
\end{table}

\begin{table}
\centering
\footnotesize
\caption{The detected architectural smells and the concurrency bugs of all projects}
\label{tab:proj:as:bugs}
\begin{tabular}{l|rrrrr} \hline 
& \textbf{Accumulo} & \textbf{Hadoop} & \textbf{Oozie} & \textbf{Spark} & \textbf{ZooKeeper} \\ \hline 
\textbf{UD}&  138 &1.132 & 36 & 39 & 128\\
\textbf{CD}& 234.449&1.249.351&  10.864 &  52.347 & 32.089\\
\textbf{HL}& 2.007 &5.421&  241 &  118 & 337\\
\textbf{MAS}& 23.6594 &1.255.904& 11.141 &  52504&  32554 \\ \hline 
\textbf{AS}& 4763.188&2.511.808 &  22.282 & 105.008 &  65.108\\ \hline 
\textbf{DATARACE}& 23 &121& 9 & 0& 66\\ 
\textbf{STARVATION}& 0 &12&   2 & 0& 3\\ 
\textbf{SUSPENSION}&18 & 33&1&2& 9\\ 
\textbf{ORDER}&2 &41&3&6& 14\\ 
\textbf{DEATHLOCK}&4 &46&  5&1& 18 \\ 
\textbf{NOTCLEAR}&0 & 16& 0 &0& 0\\ 
\textbf{ATOMICITY}& 0& 16&1&0& 12\\ 
\textbf{LIVELOCK}& 0 & 5& 0 & 0& 0\\ \hline 
\textbf{BUGS} & 47 & 290 & 21 & 9& 122\\ \hline 
\end{tabular}
\end{table}


Apache Hadoop is the biggest project, the most popular one and its developer team has a very high work activity respect to the other projects of the dataset. Apache Spark has the biggest team size of the dataset with 1552 developers. Apache Ozzie is the less popular of the dataset, but it has a medium team size with a moderate development activity. Apache Accumulo is a medium project size with a large community. Apache Zookeeper is the smallest project of the dataset and it has a small and very low active development team.

For each release of the 5 projects, the authors elicited JIRA\footnote{an issue management platform allowing users to manage their issues throughout their entire life cycle} issue-tracking information and provided a list of issues and buggy-classes per release, manually mapping every issue to a concurrency bug \cite{AsadollahSEH17}. In our study, we reuse the same dataset, manually confirming every instance in the dataset \footnote{openly available online: \url{https://goo.gl/wcdD16}} (all the compile-available versions). To understand the inter-rater agreement (IRR) of our manual coding with respect to the original dataset inherited from related work \cite{AsadollahSEH17}, we employed the well-known Krippendorff Alpha coefficient \cite{krippendorff04} --- in the scope of this study, the IRR agreement between our manual coding and the previously-coded dataset amounted to $\alpha=0,84$ which is $>>$ than the usually considered acceptable value of 0,800.

Further on, we operated over the dataset as follows.
First, we cloned every release found on the dataset. Second, we ran Arcan detection over the release. Finally, we study the correlation between the human coding of concurrency bugs available in Asadollah et al.~\cite{AsadollahSEH17} and the architecture smell detection provided by Arcan. 

Table~\ref{tab:proj:as:bugs} shows the distribution of  each concurrency bug and  architectural smell computed by Arcan with the respective total number (AS and BUGS columns). We can observe that Spark project is the project with less bugs but with more architectural smells than Oozie and ZooKeeper. Moreover, Hadoop is the project having more bugs and architectural smells than the other projects as expected, since we could consider much more versions ($76$) than other projects in the dataset.
\subsection{Data Analysis}\label{sec:performed:analysis}
In this Section we describe the data analyses protocol.
\begin{itemize}
\item \emph{Correlation Analysis}: we performed two correlation analyses using the Spearman rank correlation coefficient~\cite{spearman04} exercised between concurrency bugs and architectural smells. Spearman rank correlation coefficient is a nonparametric measure of rank correlation (statistical dependence between the rankings of two variables). It assesses how well the relationship between two variables can be described using a monotonic function.
The Spearman correlation between two variables assesses monotonic relationships between those variable (whether linear or not). If there are no repeated data values, a perfect Spearman correlation of +1 or -1 occurs when each of the variables is a perfect monotone (increasing or decreasing) function of the other.
This analysis was conducted on the complete dataset and, subsequently, focusing on each individual project. Moreover, we computed the Kendall~\cite{kendall1938new} and Pearson~\cite{pearson1895note} correlation during the setup phase of the study.

{\color{black}Pearson correlation is the most used one to measure the relationship degree between linearly related variables, while Kendall rank correlation is another of the non-parametric tests applied to measure the strength of dependence between two variables~\cite{Croux2010}.}

We have not reported those results since we noticed that the results were comparable with the Spearman correlation tests.

\item \emph{Associative-Rule Mining}: we studied the associative rules existing in the dataset using the Apriori~\cite{Agrawal:1994:FAM:645920.672836} algorithm which is defined for discovering all significant associative rules between items in a large database of transaction. However, we used the algorithm to study whether architectural smells could be associated with one or more concurrency bugs and vice versa, computed using the \texttt{apriori}\footnote{\url{https://cran.r-project.org/package=arules}} R package.
\end{itemize}


\section{Empirical Study Results}\label{sec:results}
In this section we describe the results obtained through the performed analysis outlined in Section~\ref{sec:performed:analysis}.

\subsection{Correlation Results}
Correlation analysis run over the entire dataset outlined in Table~\ref{tab:spearman:as:bug}. The table shows the correlation test results of Spearman tests among any type of architectural smell and any type of concurrency bug. Moreover, the last column of the table reports the results of the correlation between architecture smells and each individual concurrency bug; we use the indication ``Not clear" when the bug instance type was not identifiable. Test results  are represented using heat-maps and therefore assume values in the range of [$-1,+1$] (from blue to red). 
We noticed that the best results are obtained through the correlation of any AS and the presence of a bug (see last column on the right of Table~\ref{tab:spearman:as:bug}). Furthermore, we noticed no correlation between several architecture smells and \emph{Suspension} bugs (SBL). 

However, even the best result obtained in our correlation tests was lower than $0.20$ --- the value is not enough to warrant a strong correlation. In this instance, we were not able to make any conclusive observation. Hence, we conducted a project-specific analysis of correlations to understand if any of the factors influencing our sample may be also mediators in the correlations under study. 

\begin{table}[ht]
\centering

\caption{Spearman's test AS over Concurrency bugs}
\label{tab:spearman:as:bug}
\resizebox{\linewidth}{!}{
\begin{tabular}{p{0.8cm}|*{9}{p{0.6cm}}} \hline 
\textbf{AS/CB}
& DR&
SS&
SBL&
OV&
DL&
NC&
AV&
LL&
BUG \\ \hline 
CD 	& -0.06	&-0.06	&0.13	&0.07	&0.05	&0.11	&-0.12	&0.04	&0.11\\
HL 	& -0.04	&-0.04	&0.12	&0.07	&0.09	&0.06	&-0.09	&0.04	&0.14\\
MAS & -0.06	&-0.02	&0.13	&0.06	&0.05	&0.11	&-0.12	&0.04	&0.11\\
UD 	& -0.01	&0.02	&0.06	&0.07	&0.05	&0.12	&-0.05	&0.08	&0.19\\
Legend &  -1.00 & -0.75 & -0.50 & -0.25&  0.00 & 0.25 & 0.50 & 0.75  & 1.00 \\ \hline 
\multicolumn{10}{l}{NC - whether is \emph{Not Clear} which is the type of concurrency bug}\\
\multicolumn{10}{l}{BUG - whether at least a concurrency bug is present }\\
\end{tabular}
}
\end{table}

This second correlation analysis  is outlined in Table~\ref{tab:spearman:as:bug:project}. The table shows the application of correlation analysis to individual projects. 

\begin{table*}[ht]
\centering
\scriptsize
\caption{Spearman's test AS over Concurrency bugs by project}
\label{tab:spearman:as:bug:project}  
\begin{tabular}{p{1.6cm}|p{1cm}|*{9}{c}}  \hline
&
&
\multicolumn{9}{c}{\textbf{Concurrency bugs}}\\ \hline 
&
&
\multicolumn{9}{c}{\textbf{}}\\
\textbf{Projects}&
AS&
DR&
SS&
SBL&
OV&
DL&
NC&
AV&
LL&
BUG\\
 \hline
Accumulo&CD & 0.16&*&-0.33&0.39&-0.02&*&*&*&0.03\\
Accumulo&HL & 0.01&*&-0.3&0.41&0.05&*&*&*&0.03\\
Accumulo&MAS & 0.16&*&-0.33&0.39&-0.02&*&*&*&0.03\\
Accumulo&UD & 0.28&*&-0.35&0.3&-0.03&*&*&*&0.13\\
 \hline
Hadoop&CD & -0.07&0.00&0.1&0.2&0.01&0.1&-0.18&0.01&0.11\\
Hadoop&HL &  -0.02&-0.1&0.00&0.23&0.13&0.01&-0.08&0.05&0.22\\
Hadoop&MAS & -0.07&-0.01&0.1&0.2&0.01&0.1&-0.18&0.01&0.11\\
Hadoop&UD &  0.00&-0.06&-0.01&0.17&0.04&0.04&-0.07&0.06&0.17\\
 \hline
Oozie&CD & 0.17&-0.41&0.15&-0.11&0.22&*&0.15&*&0.24\\
Oozie&HL &  0.2&-0.55&0.15&-0.11&0.22&*&0.15&*&0.18\\
Oozie&MAS & 0.17&-0.41&0.15&-0.11&0.22&*&0.15&*&0.24\\
Oozie&UD &  -0.2&0.45&-0.15&0.19&-0.22&*&-0.15&*&-0.22\\
 \hline
Spark&CD & *&*&0.42&0.03&-0.28&*&*&*&0.17\\
Spark&HL & *&*&0.24&0.37&-0.33&*&*&*&0.36\\
Spark&MAS & *&*&0.42&0.03&-0.28&*&*&*&0.17\\
Spark&UD & *&*&0.38&-0.34&0.38&*&*&*&0.06\\
 \hline
Zookeeper&CD & -0.07&-0.06&0.16&-0.14&-0.04&*&-0.02&*&-0.17\\
Zookeeper&HL &  -0.06&-0.01&0.21&-0.17&-0.07&*&-0.04&*&-0.13\\
Zookeeper&MAS & -0.07&-0.06&0.17&-0.15&-0.03&*&-0.03&*&-0.17\\
Zookeeper&UD &  0.01&0.13&0.12&-0.16&-0.1&*&0.02&*&-0.03\\
 \hline
\multicolumn{2}{c}{Legend} &  -1.00 & -0.75 & -0.50 & -0.25&  0.00 & 0.25 & 0.50 & 0.75  & 1.00 \\
\multicolumn{11}{c} {}\\
\multicolumn{11}{l}{NC - whether is \emph{Not Clear} which is the type of concurrency bug}\\
\multicolumn{11}{l}{BUG - whether at least a concurrency bug is present }\\
\multicolumn{11}{l}{\textbf{*} correlation non computed since there are not bug of the specified type in that projects}\\
\multicolumn{11}{l}{or the variance was $0$ or the associated p-value was $> 0.05$}\\ \hline
\end{tabular}
\end{table*}

We observe that Data-Race (DR) seems meagerly correlated with all types of architecture smells with a peak correlation value of $0.28$ with UD in the scope of the Accumulo project, an intrinsically distributed datastore based on the Google BigTable model. 
In addition, we observe that the Starvation (SS) bug shows the highest correlation values in Oozie project where the values showed a negative monotone correlation with CD, HL, and MAS architectural smells assuming values $-0.41$, $-0.55$ and $-0.41$ respectively; there is a positive monotone correlation with UD with value $0.45$. This correlation behavior seems to indicate that  Starvation bugs are growing in number when UD grow linearly and while other AS are decreasing at the same time.

Suspension bug (SBL) has a negative correlation with all AS of at least $-0.3$ in Accumulo project but exhibits a positive correlation in Spark; in particular, the bug has a correlation value of $0.42$ with CD and MAS, $0.38$ with UD and $0.24$ with HL. This seems to indicate that the Suspension bug decreases while the AS are increasing in Accumulo. However, the Suspension bug increases while the AS are decreasing in Spark. This evidence leads to the conclusion that this type of bug is intrinsically dependent on the project characteristics similarly to Data-Race conditions with the type of the project, but could be related to the developers since more developers are involved, more likely the correlation increase (Spark) or decrease (Accumulo) with less developers involved.

Order violation (OV) bugs have the highest positive correlation with all AS in Accumulo, and also in Hadoop and Spark. This seems to indicate that architecture smells may cause nasty Order violation circumstances in highly distributed systems such as in the projects listed above.

Deadlock bugs (DL) have a positive correlation in Oozie for CD, HL and MAS with a value of $0.22$; there is a negative correlation in Spark with values $-0.28$, $-0.33$ and $-0.28$ respectively. Moreover, UD has a negative correlation in Oozie  of $-0.22$ (positive with the other AS) and a positive correlation in Accumulo of $0.38$ (negative with the other AS). This trend may be connected to project Oozie's nature of a centralized work-flow manager --- in this instance, it could mean that the more centralized the architectural structure, the more prone to deadlocks is.

Finally, concerning the remaining concurrency bugs we explored in the scope of our dataset, namely, the Notclear (NC) indicator, Atomicity (AV) and Livelock (LL) bug, we observed no strong correlations in any project as outlined in the results obtained  on the full dataset and showed in Table~\ref{tab:spearman:as:bug}. For these bugs, there may \emph{not} exist any form of structural/behavioral relation.

\begin{framed}
\textbf{Findings.}
The relation between architecture smells and concurrency bugs is non-trivial and runs deeper than the structural correlation analysis we operated in the scope of this paper. Nevertheless, we observed interesting and project-specific behaviors for Data-Race, Suspension, Order violation, and Deadlock conditions --- their architectural implications deserve further study. Finally, our observations concerning remaining bugs is inconclusive.
\end{framed}

\subsection{Associative Rules Results}
\begin{figure}[]
\centering
\includegraphics[width=0.8\linewidth, height=7.5cm]{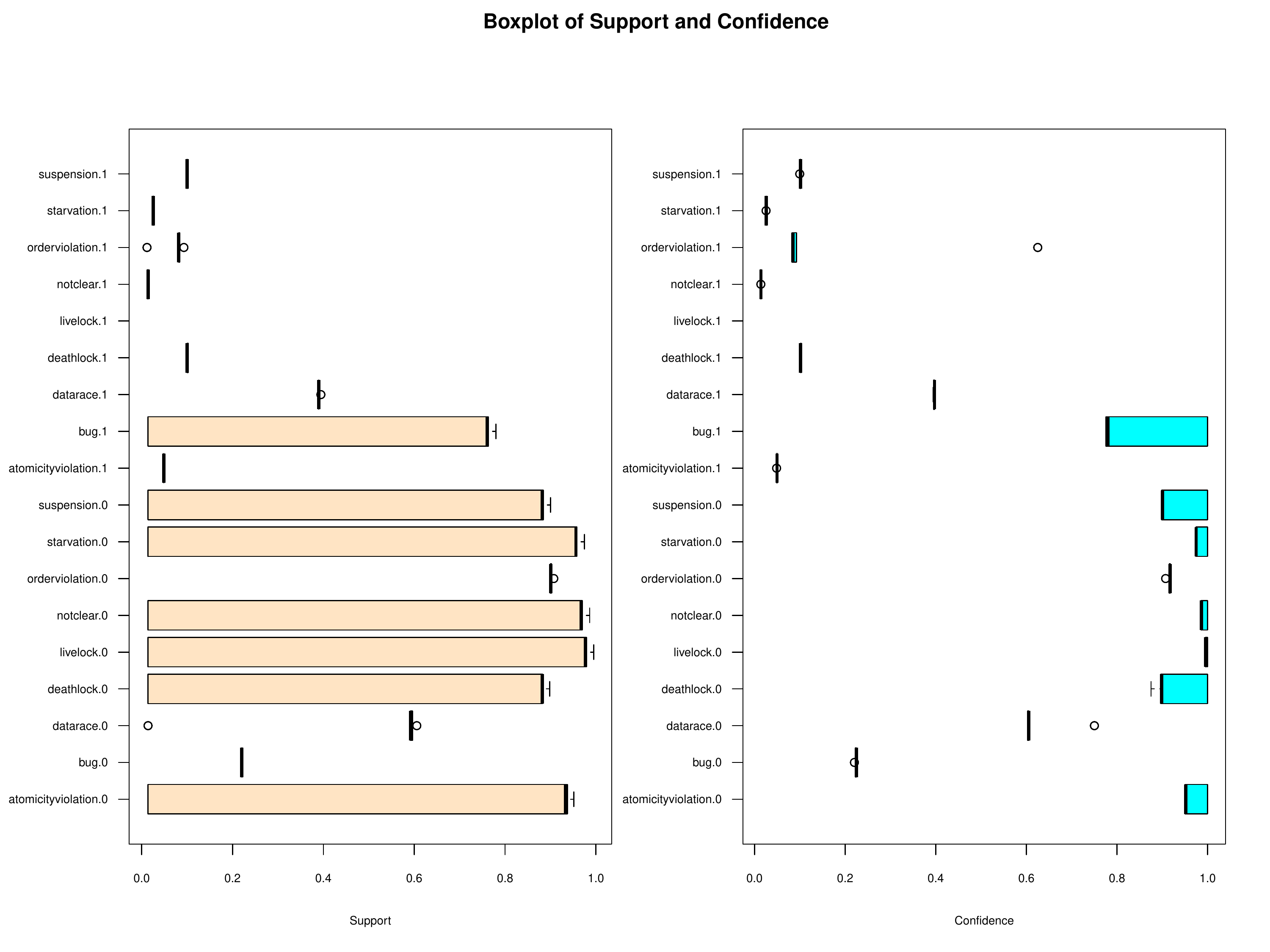}
\caption{Boxplot of Support and Confidence of Apriori rules among Bugs. The presence of the BUG type is given by the value $1$ and the absence is given by the value $0$.}
\label{fig:boxplot:rules:bug}
\end{figure}

\begin{figure}[]
\centering
\includegraphics[width=0.8\linewidth, height=7.5cm]{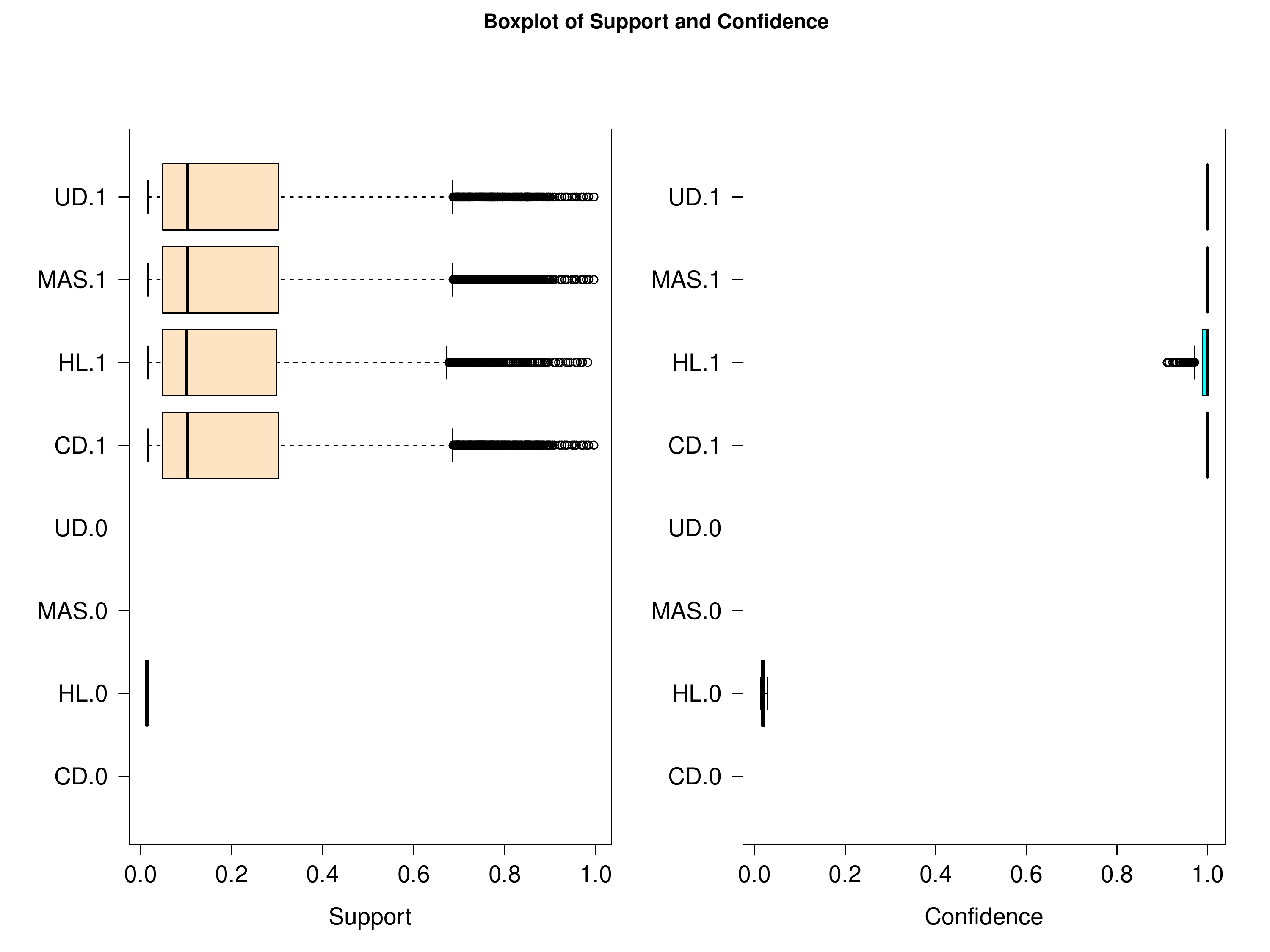}
\caption{Boxplot of Support and Confidence of associative rules among AS. The presence of the AS is given by the value $1$ and the absence is given by the value $0$.}
\label{fig:boxplot:rules:as}
\end{figure}

In this section we describe the obtained results of the associative rules from the execution of the Apriori association rule mining algorithm.
Figure~\ref{fig:boxplot:rules:bug} and Figure~\ref{fig:boxplot:rules:as} show the \emph{support} and \emph{confidence} of the extracted associative rules to identify whether an AS type is associated to a bug and vice versa. 
Moreover, Figure~\ref{fig:boxplot:rules:bug} and Figure~\ref{fig:boxplot:rules:as} show both the presence and the absence of a bug or AS:
\begin{itemize}
\item the \emph{presence} of a bug or an AS is given by indicating the value 1 (e.g., $AS.1$, $bug.1$);
\item the \emph{absence} of a bug or an AS is given by showing the value 0 (e.g., $AS.0$, $bug.0$).
\end{itemize}

We start explaining the associative rules of a presence or absence of a bug given any AS. Figure~\ref{fig:boxplot:rules:bug} shows that the confidence and support are too low for the learned associative rules related to the presence of a specific type of bug. However, several associative rules related to $bug.1$ have a support and confidence higher than $0.7$. Most relevant rules associated to the presence of AS are summarized in Table~\ref{tab:rules:bugs:as}. 
 The table shows the rules related to the presence of any given concurrency \emph{bug};  AS results are grouped by \emph{lift}, using a cut-off value of $0.7$ in both support and confidence. The rules without HL and UD architecture smells, obtained better results --- this uncovers a clear relation existing among concurrency bugs, CD, and MAS. Furthermore, we observed that the HL rules score slightly better results than UD rules.

\begin{table}[htbp]
  \centering

  \caption{Rules of AS vs. a generic Bugs}
\resizebox{\linewidth}{!}{
\begin{tabular}{r|l|l|l}
    \hline 
    \textbf{Rules} & \textbf{Support} & \textbf{Confidence} & \textbf{Lift} \\
    \hline 
    \{$cd=1$\} $\Rightarrow$ \{$bug=1$\} & 0.779 & 0.779 & 1 \\
    \{$mas=1$\} $\Rightarrow$ \{$bug=1$\} & 0.779 & 0.779 & 1 \\
    \{$cd=1,mas=1$\} $\Rightarrow$ \{$bug=1$\} & 0.779 & 0.779 & 1 \\
    \hline 
    \{$hl=1$\} $\Rightarrow$ \{$bug=1$\} & 0.766 & 0.776 & 0.996 \\
    \{$cd=1,hl=1$\} $\Rightarrow$ \{$bug=1$\} & 0.766 & 0.776 & 0.996 \\
    \{$hl=1,mas=1$\} $\Rightarrow$ \{$bug=1$\} & 0.766 & 0.776 & 0.996 \\
    \{$cd=1,hl=1,mas=1$\} $\Rightarrow$ \{$bug=1$\} & 0.766 & 0.776 & 0.996 \\
    \hline 
    \{$ud=1$\} $\Rightarrow$ \{$bug=1$\} & 0.761 & 0.775 & 0.995 \\
    \{$hl=1,ud=1$\} $\Rightarrow$ \{$bug=1$\} & 0.761 & 0.775 & 0.995 \\
    \{$cd=1,ud=1$\} $\Rightarrow$ \{$bug=1$\} & 0.761 & 0.775 & 0.995 \\
    \{$ud=1,mas=1$\} $\Rightarrow$ \{$bug=1$\} & 0.761 & 0.775 & 0.995 \\
    \{$cd=1,hl=1,ud=1$\} $\Rightarrow$ \{$bug=1$\} & 0.761 & 0.775 & 0.995 \\
    \{$hl=1,ud=1,mas=1$\} $\Rightarrow$ \{$bug=1$\} & 0.761 & 0.775 & 0.995 \\
    \{$cd=1,ud=1,mas=1$\} $\Rightarrow$ \{$bug=1$\} & 0.761 & 0.775 & 0.995 \\
    \{$cd=1,hl=1,ud=1,mas=1$\} $\Rightarrow$ \{$bug=1$\} & 0.761 & 0.775 & 0.995 \\
    \hline 
    \end{tabular}%
    }
  \label{tab:rules:bugs:as}%
\end{table}%

For what concerns associative rules constructed by using concurrency bugs to assess the absence and presence of AS,  Figure~\ref{fig:boxplot:rules:as} shows that the obtained rules have high confidence to indicate the presence of AS types, but the box of the boxplot is condensed to low values of the support. The rules extracted for the absence of architectural smells are not relevant (lower than $0.2$).
Figure~\ref{fig:boxplot:rules:as} shows that the confidence and the support of the rules is low in the case of AS absence (values lower than $0.2$). However, the confidence for the presence of AS is higher than $0.8$ in most all the cases (e.g., UD.1, MAS.1 and CD.1) and the support have many outliers higher than $0.8$. We consider bugs as good predictors for a general AS, as shown at Figure~\ref{fig:boxplot:rules:as}. We present some of the rules related to the presence of bugs in Table~\ref{tab:rules:as:bug}. 
Table~\ref{tab:rules:as:bug} shows the rules associated to each type of architectural smell given by  bugs. A bug is associated with a Support equal to or higher than $0.98$ and a Confidence equal to or higher than $0.76$ to the AS. 

\begin{table}
\centering
\footnotesize
\caption{Associative rules related to AS}
\label{tab:rules:as:bug}
\begin{tabular}{l|l|l|l}
    \hline 
\textbf{Rules}	&	\textbf{Support}	&	\textbf{Confidence}	&	\textbf{Lift}	\\
    \hline 
\{$bug=1$\} $\Rightarrow$ \{$cd=1$\}	&	0.78	&	1.00	&	1.00	\\
\{$bug=1$\} $\Rightarrow$ \{$mas=1$\}	&	0.78	&	1.00	&	1.00	\\
\{$bug=1$\} $\Rightarrow$ \{$hl=1$\}	&	0.77	&	0.98	&	1.00	\\
    \hline 
\{$bug=1$\} $\Rightarrow$ \{$ud=1$\}	&	0.76	&	0.98	&	0.99	\\
    \hline 
\end{tabular}
\end{table}

Moreover, the extraction of associative rules per projects was performed. We identified the same rules among the projects. Support and Confidence of the associative rules extracted from the projects with less bugs (i.e. Accumulo, Oozie and Spark) have slightly higher values than the other projects (i.e. Hadoop and Zookeeper). In conclusion,  the results obtained in general on all the projects are still valid also for single projects.

\begin{framed}
\textbf{Findings.}
The associative rules of the absence and presence of a concurrency bug and one (or more) architectural smell(s) show that CD and MAS give the best results associated to the presence of a concurrency bug and vice versa. The absence of a concurrency bug is not associated to the presence of an architectural smell and similarly the absence of an architectural smell is not associated to the presence of one (or more) concurrency bug(s).
\end{framed}

\section{Discussions}
In order to answer our RQ$_1$, we performed Spearman's correlation tests by project and for the entire projects as explained in Section~\ref{sec:results}. We noticed no relevant correlation at dataset level, since the values are lower or equal to $0.18$. We studied the correlation per project between concurrency bugs and architectural smells, since we argue that it is in fact possible to highlight correlations with respect to smells only in the context of some specific project characteristics. For example, Suspension-based locking bugs show a negative correlation with all AS in the Accumulo project, but positive with all other projects and significantly stronger with Spark. In contrast with the other projects, we observe that Accumulo teams are prone to reducing the Unstable Dependency architecture smell while they are more prone on introducing new Suspension-based locking bugs. In addition, Starvation has a negative correlation with all the AS in the Arcan detection set and its presence is predominant in the Oozie project. 

Answering to RQ$_2$, the presence of a concurrency bug is not correlated in general to any AS across most projects, as shown in the \emph{BUG} column from Table~\ref{tab:rules:as:bug}.
To offer a more focused lens of analysis over the influences we investigated, we extracted the associative rules using the Apriori algorithm, as explained in Section~\ref{sec:results}. We observed that, while it is hard to discover associative rules between architecture smells and a particular type of concurrency bug, it is in fact evident that any AS is associated to the occurrence of a concurrency bug with a support and confidence higher than or equal to $0.76$. Moreover, a type of concurrency bug is related to any AS where we extracted rules with a support and confidence higher than or equal to $0.76$.

We conclude that there are in fact relations and implications between software architecture smells and concurrency bugs, but the theory we outlined in Sec. 3.4 and all the relations thereto need further exploratory research to pinpoint any characteristics that link concurrency bugs to architecture smells further --- in the scope of this paper we have been able to show that a relation does in fact exist and is considerable enough to warrant further exploration. 

\section{Threats to Validity}\label{sec:validity}


\textbf{External Validity} By means of an externally-coded dataset which was prepared previously to our study and in a totally independent fashion, we aimed explicitly to strengthen external validity of this study. 
We explicitly aimed at controlling variables that could warrant a higher external validity, as the variables we controlled in the several options available to us for dataset reuse. 
Although a number of variables remain uncontrolled (e.g., programming language, code structure complexity, PL type, etc.) due to tool or other limitations.
{\color{black} Arcan detection performance was evaluated in two industrial case studies based on the feedback of the developers is described in~\cite{Arcelli2017-ICSA}, where the authors report a precision of 100\%, since Arcan found only correct instances of architectural smells, and a 63\% value for recall. The developers reported five more architectural smells, which were false negatives, related to 180 external components outside the tool's scope of analysis.
Moreover, in order to increase the reliability on Arcan, a manually validation was performed in ten open-source projects~\cite{Arcelli2017-ICSA} and in four industry projects on which the practitioners evaluated the usefulness of the tool support ~\cite{Martini2018}. 
The detection tool usefulness was evaluated considering practitioners feedbacks on the the automatic detection support of architectural smells detected by Arcan~\cite{Martini2018}.}

\textbf{Construct} and \textbf{Internal Validity}. The methods we adopted to evaluate our solution are less prone to generalizability issues since their evaluation re-uses established content analysis techniques and 
correlation analysis previously known and widely-used in the literature. Given that the comparison we operate in the context of this study relates to evaluating agreement with the assessment made by independent human observers (i.e., the original labelers of concurrency bugs in our dataset), we chose to run a simple Spearman correlation test evaluation. We used a library implemented in \emph{R} called \textit{Corrplot}\footnote{\url{https://cran.r-project.org/package=corrplot}}.
Other correlation tests exist, e.g., Pearson and Kendall and we test them in the setup study phase anyway. 
{\color{black} We considered only four architectural smells since only few available tools are able to detect several architectural smells compared to tools to detect code debt.}

 

\textbf{Conclusion Validity}. Our conclusions are based on the manual inspection of the correlation tests scores between a human predictor triangulated independently in a completely unrelated study and the precision of the Arcan, architecture smells detector. We argue that our interpretation is valid since it is limited to assessing whether the overlap successfully addresses our research question.


\textbf{Replication Package}. We provide an extensive replication package containing: 
(a) the version of Arcan that was used for the purpose of this study, as well as instructions to setup and run the tool\footnote{Arcan and instruction for setup and run the tool, \url{https://goo.gl/DhwHPq}}; 
(b) the original dataset from Asadollah et Al.
\footnote{Dataset of Concurrency bugs, \url{https://goo.gl/DFg5f5}};
 (c) The data extracted by Arcan\footnote{Dataset of Arcan, \url{https://goo.gl/Jw51ND}}. 
\\

\section{Conclusions and Future Work}\label{sec:conc}

This paper seeks to shed light over the possible  relations between software architecture smells and concurrency bugs. In the context of our study, we used Arcan, a configurable tool to detect and rank dependency-based architecture smells. More in particular, we employed Arcan on 125 releases of 5 large open-source products. We analyzed Arcan results in combination with the concurrency bugs reported independently across the dataset by external observers.
The relation between architecture smells and concurrency bugs is non-trivial and runs deeper than the structural correlation analysis we operated in the scope of this paper. Nevertheless, we observed interesting and project-specific behaviors for Data-Race, Suspension, Order violation, and Deadlock conditions.
Finally, our observations concerning remaining bugs are inconclusive or may suggest no relation with structural/behavioral property.
As a major contribution of this study, we assessed that there is in fact an empirical relation between architectural smells and concurrency bugs. Namely, our study reveals that there is in fact a sensible relation between architecture smells and several Deadlock, and Suspension conditions. These relations need further, more specific and possibly qualitative exploratory research asking developers feedbacks, but provide significant hints  to developers on parts to be carefully checked. In the future, we plan to investigate more thoroughly the role of each individual architecture smell in the scope of each individual concurrency bug we are targeting in the scope of this study.
Moreover, we plan to instrument a case-study campaign specifically designed to further understand the architectural implications of every single concurrency bug, from a combined qualitative and quantitative perspective.
We could also explore the impact of the severity of the architectural smells and the Architectural Debt Index~\cite{Roveda2018}.
We have considered the architectural smells detected by Arcan, but others can be considered in the future not strictly based on dependency issues, as for example the micro services smells \cite{Taibi2018}. 


\bibliographystyle{spbasic}   
\bibliography{arch,essere,bibliounica}

\end{document}